\documentclass[aps,10pt,prd,twocolumn,nofootinbib]{revtex4-2}
\usepackage{bm,dcolumn,amsmath,graphicx,amsfonts,amssymb,physics}
\usepackage{tablefootnote}
\usepackage{hyperref,xcolor}
\definecolor{cite}{rgb}{0.,0.,0.9}
\hypersetup{colorlinks,linkcolor={cite},citecolor={cite},urlcolor={cite}}

\renewcommand{\v}[1]{\ensuremath{\boldsymbol{#1}}}
\def\d{\ensuremath{{\rm d}}}

\newcommand{\matel}[3]{\langle{#1}|{#2}|{#3}\rangle}
\newcommand{\redmatel}[3]{\langle{#1}\|{#2}\|{#3}\rangle}

\begin{document}

\title{Atomic structure calculations for constraining new electron-electron forces}

\date{28 October 2026} 

\author{Cameron J. West}

\author{Benjamin M. Roberts}\email[]{b.roberts@uq.edu.au}

\affiliation{School of Mathematics and Physics, The University of Queensland, Brisbane QLD 4072, Australia}

\begin{abstract}
\noindent

We present a general approach for calculating new electron-electron interactions with atomic structure theory, include dominant many-body effects, and extract bounds for parity-odd cases. The procedure is valid for any mediator mass and improves previous calculations by including many-body corrections to the new interaction through the time-dependent Hartree-Fock method. Bounds on coupling strengths are then extracted for scalar-pseudoscalar and vector-axial vector interactions by comparison with experimental atomic electric dipole moments and parity non-conserving (PNC) transitions respectively. In the latter case, our calculations rule out any coupling-mass ratio $|g^e_{\mathrm{V}}g^e_{\mathrm{A}}|/m_X^2$ greater than $10^{-11}\,\mathrm{MeV^{-2}}$ above $m_X\sim10\,{\rm MeV}$, covering a previously unconstrained mass region. We also report an updated calculation of the regular (Standard Model) electron-electron $Z^0$ exchange contribution to the $6s$-$7s$ PNC transition amplitude $E_{\mathrm{PNC}}$ in cesium.

\end{abstract}

\maketitle

\section{Introduction}
Anomalous observations for which the Standard Model fails to account, such as the dark matter \cite{bertone_particle_2005, safronova_search_2018} and matter-antimatter asymmetry \cite{sakharov_violation_1967, dine_origin_2003} problems, continue to motivate searches for its extension. New leptonic interactions are one possibility \cite{safronova_search_2018, buras_global_2021, cong_spindependent_2025, eberhart_leptophilic_2025} which can be constrained by comparing theory with experiment and, through their mediators, contribute to broader searches for new particles. Precision atomic physics can spectroscopically probe electron-electron ($ee$) and electron-nucleus ($eN$) interactions, although there is a scarcity in investigations of the more complex two-body $ee$ case, especially those which violate parity ($\mathcal{P}$) \cite{safronova_search_2018, cong_spindependent_2025}. In light of this, we provide a general theoretical framework for performing atomic structure calculations in the presence of $ee$ interactions and demonstrate their utility by extracting and reporting new bounds on their mediators' coupling constants.

Calculations of new $ee$ interactions are typically made using an approximation of the mediator's mass $m$ \cite{stadnik_improved_2018, maison_axionmediated_2021, maison_static_2022, prosnyak_updated_2023}. However, for most systems, neither the massless ($m\rightarrow0$) nor contact ($m\rightarrow\infty$) limits are valid in the important keV to MeV region where, like the high-mass case, fewer constraints exist \cite{cong_spindependent_2025}. We derive a new expression for the interaction potentials which is valid in all regions, enabling accurate calculations for all mediator masses. This  approach reduces to the limiting values as appropriate and discerns important features in the intermediate mass range, making it a robust improvement on existing methods.

New interactions introduce a potential to the atomic Hamiltonian. We describe it in general and provide expressions for its inclusion in many-body Dirac-Hartree-Fock (DHF) atomic structure calculations. The time-dependent Hartree-Fock (TDHF) equations \cite{dirac_note_1930, dalgarno_timedependent_1966, johnson_relativistic_1980, dzuba_relativistic_1984a, dreuw_singlereference_2005} are then applied to account for first-order corrections induced by the new $ee$ interaction to the DHF potential and wavefunctions. This is equivalent to the random phase approximation (RPA)  \cite{johnson_relativistic_1980}, and is necessary for $\mathcal{P}$-violating interactions if normal symmetry assumptions for the DHF Hamiltonian are to be preserved. Significant higher order electron correlation effects are included by using the all-order correlation potential method \cite{dzuba_screening_1988, dzuba_summation_1989, roberts_electricdipole_2023}. We predominantly consider single-valence systems but also demonstrate extension to multi-valence atoms with the combined configuration interaction and many-body perturbation theory  (CI+MBPT) method \cite{dzuba_combination_1996}.

The new potential is valid for all possible electron-electron interactions mediated by scalar ($S=0$) or vector ($S=1$) bosons and induces parity-dependent atomic effects. For example, $\mathcal{P}$-even potentials yield energy shifts, while $\mathcal{P}$-odd mix states of opposite parity inducing otherwise forbidden matrix elements of atomic operators. For the electric dipole operator, this corresponds to atomic electric dipole moments (EDMs) and parity non-conserving (PNC) transitions. While our formalism applies to all cases, we restrict our calculations to these, as $\mathcal{P}$-even interactions are already well constrained from light systems \cite{cong_spindependent_2025}.

Scalar-pseudoscalar interactions ($\mathcal{P}$-odd, $S=0$) are a promising example of new physics because their mediators (axion-like particles) are popular dark-matter candidates \cite{preskill_cosmology_1983, abbott_cosmological_1983, dine_notsoharmless_1983, safronova_search_2018, eberhart_leptophilic_2025} which, as a source of $\mathcal{P}$, $\mathcal{T}$-violation, may also contribute to explanations of the universe's matter-antimatter asymmetry \cite{sakharov_violation_1967}. They can be unambiguously constrained by their induced atomic EDMs since these are otherwise significantly suppressed within the Standard Model \cite{pospelov_electric_2026}. We perform calculations to extract bounds on the interaction from experimental EDMs in $^{133}$Cs \cite{murthy_new_1989} and $^{205}$Tl \cite{regan_new_2002}, producing limits on the interaction's coupling strength for all mediator masses. While our new bounds are less stringent than existing diamagnetic and molecular limits \cite{stadnik_improved_2018, maison_axionmediated_2021, maison_static_2022, prosnyak_updated_2023}, they account for important many-body effects and reveal areas of insensitivity to new bosons with intermediate masses.

We also perform vector-axial vector ($S=1$, $\mathcal{P}$-odd) calculations and report new bounds for previously unconstrained mediator masses. Instead of EDMs, these interactions mediate $E$1-forbidden transitions, such as $E_{\mathrm{PNC}}^{6s\rightarrow7s}$ in cesium, for which the $ee$ contribution has received less attention than the dominant $eN$ interaction \cite{dzuba_relativistic_1984, dzuba_relativistic_1985,blundell_highaccuracy_1990, blundell_highaccuracy_1992,  dzuba_highprecision_2002, flambaum_radiative_2005, porsev_precision_2009, dzuba_revisiting_2012, sahoo_new_2021, roberts_comment_2022, dzuba_longrange_2022}. First, we calculate the Standard Model $ee$ interaction due to $Z^0$ exchange, which is in reasonable agreement with previous values \cite{blundell_highaccuracy_1990, blundell_highaccuracy_1992} and includes an improved many-body contribution. Then, we generalise to other masses within the contact limit to extract new constraints on the interaction strength of an arbitrary leptophilic vector boson by comparison with existing measurements of $E_{\mathrm{PNC}}^{6s\rightarrow7s}$ \cite{wood_measurement_1997}. These are the first direct bounds obtained outside of the ultralight region\footnote{Other bounds have been obtained indirectly by combining astrophysical and laboratory data, see Ref.~\cite{cong_spindependent_2025} and references therein for discussions of such cases.}.

Finally, we leave calculations of $\mathcal{P}$-even cases, which induce energy shifts, for future work. These are most relevant to light systems where many-body effects are small enough to discern such new physics signals; Ref.~\cite{cong_spindependent_2025} provides a comprehensive review. More recent examples include constraints obtained from helium \cite{cong_testing_2026} through ionisation energy discrepancies \cite{patkos_complete_2021,clausen_ionization_2021, clausen_ionization_2025, clausen_metrology_2025, bondy_theory_2025} and in lithium-like systems \cite{abdullin_axionexchange_2026} by comparison with transition levels. Calculations similar to those formalised here have ruled out all but a low mass ($m < 800$ eV) scalar-scalar contribution to He discrepancies and extracted new bounds for an axial vector-axial vector interaction from $^{209}$Bi$^{80+}$. These may also be relevant to energy-sensitive tests in heavy elements, such as isotope shifts, which have already been employed as a probe for new physics \cite{berengut_precision_2025}.

\section{Theory}\label{sec:theory}
Fermionic interactions can be described by the effective interaction Lagrangian
\begin{align}
    \mathcal{L}_{\mathrm{int}} &= -\hbar c a\sum_{\psi,\alpha}\bar{\psi}G_\alpha\psi \label{eq:lag},
\end{align}
where $a$ is the mediating boson, $\bar{\psi}=\psi^\dag\gamma_0$, and the sum is over all fermions $\psi$ and possible boson-fermion interactions $\alpha$, with $\alpha=\{\mathrm{S},\mathrm{P}\}$ (scalar, pseudoscalar) for a scalar boson (where $a=\phi$) and $\alpha=\{\mathrm{V},\mathrm{A}\}$ (vector, axial-vector) for a vector boson (where $a=A_\mu$). The interaction-specific contributions $G_\alpha$ are $G_{\mathrm{S}} =  g^e_{\mathrm{S}}$, $G_{\mathrm{P}} = -ig^e_{\mathrm{P}}\gamma^5$, $G_{\mathrm{V}} = g^e_{\mathrm{V}} \gamma^\mu$ and $G_{\mathrm{A}} = -g^e_{\mathrm{A}} \gamma^\mu \gamma^5$, where $g_\alpha$ are dimensionless coupling constants characterising the interaction strength and $\gamma^\mu, \gamma^5$ are Dirac matrices. Electron-electron interactions consist of two boson-fermion vertices and are summarised in Table~\ref{tab:interactions}. 

\begin{table}[ht]
    \centering
    \caption{Summary of electron-electron interactions $\alpha\beta$. The two boson-electron contributions are identified by $\alpha$ and $\beta$, $\Gamma^{\alpha\beta}_{jk}$ encodes their effect on the resulting potential [Eq. (\ref{eq:yuk})], and $\mathcal{T}^{\alpha\beta}_{jk}$ their effect on electron orbitals' radial components [Eq. (\ref{eq:B})]. Dirac matrices acting on electron $j$ are denoted by $\gamma^0_j$, $(\gamma^0\gamma^\mu)_j$, and $\gamma^5_j$, and generalised Pauli matrices, $\tau\in\mathbb{C}^{2\times2}$, are given numerically by  $\tau^1=\sigma_x$, $\tau^2=\sigma_y$ and $\tau^3=\sigma_z$.}
    \label{tab:interactions}
    \begin{ruledtabular}
    \begin{tabular}{ccccc}
    $\alpha\beta$ & Spin & Parity & $\Gamma^{\alpha\beta}_{jk}$ & $\mathcal{T}^{\alpha\beta}_{jk}$ \\\hline
         SS & $0$ & $+$ & $1$ & $\tau^3_j\tau^3_k$ \\
         PP & 0 & $+$ & $-\gamma^5_j\gamma^5_k$ & $\tau^1_j\tau^1_k$ \\
         SP  & $0$ & $-$ & $i(\gamma^5_j+\gamma^5_k)$ & $-\tau^3_j\tau^1_k$ \\\hline
         VV & $1$ & $+$ & $(\gamma^\mu)_j(\gamma_\mu)_k
         $ & $-I_jI_k$\\
         AA & $1$ & $+$ & $(\gamma^\mu\gamma^5)_j(\gamma_\mu\gamma^5)_k
         $ & $\tau^2_j\tau^2_k$ \\
         VA & $1$ & $-$ & $(\gamma^\mu)_j(\gamma_\mu)_k
         (\gamma^5_j+\gamma^5_k)$ & $I_j\tau^2_k$ 
    \end{tabular}
    \end{ruledtabular}
\end{table}

New interactions which satisfy Eq.~(\ref{eq:lag}) introduce an additional two-body potential $V^{\alpha\beta}$ into the atomic Hamiltonian. Factoring out the unknown coupling constants by defining $V^{\alpha\beta}=g^e_\alpha g^e_\beta V_0^{\alpha\beta}$, it takes the form of a Yukawa potential, 
\begin{align}\label{eq:yuk}
    V^{\alpha\beta}_0 &= \frac{\hbar c}{8\pi}\sum_{j\neq k}\frac{e^{-\mu r_{jk}}}{r_{jk}}\gamma^0_j\gamma^0_k\Gamma^{\alpha\beta}_{jk},
\end{align}
where the sum is over all electrons $j\neq k$, $\mu$ is related to the boson's $\mu = mc/\hbar$, $r_{jk} = |\v{r}_j-\v{r}_k|$ and $\Gamma^{\alpha\beta}_{jk}$ contains the interaction-specific terms described in Table~\ref{tab:interactions}.

The energy shifts induced by $\mathcal{P}$-even interactions ($\alpha=\beta$) can be calculated by evaluating $\matel{\Psi}{V_0^{\alpha\alpha}}{\Psi}$, where $\Psi$ is an atomic wavefunction. Comparison with existing energy discrepancies and uncertainties $\delta E$ yields upper bounds on interaction strengths,
\begin{align}
    g^e_\alpha g^e_\alpha \le \frac{\delta E}{\matel{\Psi}{V_0^{\alpha\alpha}}{\Psi}}.
\end{align}

As $\mathcal{P}$-odd interactions ($\alpha\neq\beta$) do not induce energy shifts, only off-diagonal matrix elements are non-zero. These mix states of opposite $\mathcal{P}$ enabling otherwise forbidden matrix elements from Standard Model operators $T$. To calculate them, the potential can be modelled as a small perturbation to atomic wavefunctions, wherein a state $\Psi$ with parity $\pm$ becomes $\widetilde{\Psi}_\pm=\Psi_\pm+\delta\Psi_\mp$, where $\delta\Psi_\mp$ is the perturbation to the wavefunction due to the parity odd potential. Matrix elements of $T$ in the presence of the new interaction can then be calculated using the sum-over-states approach,
\begin{align}
    T^{\alpha\beta}_{wv}\equiv\matel{\widetilde{\Psi}_w}{T}{\widetilde{\Psi}_v}_{\alpha\beta} &= \sum_{n\neq v}\frac{\matel{\Psi_w}{T}{\Psi_n}\matel{\Psi_n}{V_0^{\alpha\beta}}{\Psi_v}}{E_v-E_n}\nonumber\\ &+ \sum_{n\neq w}\frac{\matel{\Psi_w}{(V_0^{\alpha\beta})^\dag}{\Psi_n}\matel{\Psi_n}{T}{\Psi_v}}{E_w-E_n}, \label{eq:wOv}
\end{align}
where the sums are over complete orthonormal basis sets $\{n\}$ and $E_a$ is the energy of an atomic wavefunction $\Psi_a$. Here, we will exclusively consider matrix elements of the electric dipole operator $T\rightarrow \v{D}=-|e|\sum_i\v{r}_i$, which sums over electrons $i$.

The SP interaction will induce an atomic EDM. For atomic wavefunctions with valence state $v$ in such an interaction this is given by the expectation value of the electric dipole operator,  $D^{\mathrm{SP}}_{vv}$. Comparison with existing experimental limits $D^{\mathrm{expt}}_{vv}$ (where no statistically significant signal is found) places constrains on the interaction couplings:
\begin{align}
    |g^e_{\mathrm{S}}g^e_{\mathrm{P}}| \le \left|\frac{D^{\mathrm{expt}}_{vv}}{D_{vv}^{\mathrm{SP}}}\right|.
\end{align}

Calculations of spin-1 interactions are complicated by the spatial terms in $\Gamma^{\alpha\beta}_{jk}$, which, in a complete treatment, should be included similarly to the Breit interaction \cite{breit_effect_1929,breit_fine_1930,breit_diracs_1932} (see also Refs.~\cite{mann_breit_1971, johnson_atomic_2007}). For the $\mathcal{P}$-even cases the Breit equivalence is exact as these induce energy shifts, whereas for the VA term the opposite parity admixture alters the behaviour of the spatial contribution. Importantly, this mixing invalidates the assumption that the space-like contribution to $\mathcal{P}$-odd interactions is small, and so the time-like terms alone are not a sufficient approximation for the interaction.

However, as calculated in Refs.~\cite{blundell_highaccuracy_1990, blundell_highaccuracy_1992} for the weak interaction, a Fierz identity in the contact limit allows this issue to be circumvented. We restrict ourselves to this mass region and leave the Breit-like spatial contribution for future work. When $\mu\rightarrow\infty$, the Yukawa potential simplifies:
 \begin{align}
     \frac{e^{-\mu r_{jk}}}{r_{jk}} \rightarrow \frac{4\pi}{\mu^2}\delta^3(r_{jk}),
 \end{align}
 where $\delta^3$ is the 3D Dirac delta. This greatly simplifies the integrals contributing to matrix elements, as will be shown in Sec. \ref{sec:me}, such that the spatial terms can then be neglected \cite{blundell_highaccuracy_1990, blundell_highaccuracy_1992}.

The VA potential induces electric dipole transitions between states, $D_{wv}^{\mathrm{VA}}$. For the weak interaction this introduces an $ee$ contribution to induced transitions, e.g. $6s\rightarrow7s$ in $^{133}$Cs. The $Z$-boson's coupling constant,
\begin{align}
    (g^e_\mathrm{V}g^e_\mathrm{A})_Z &= \frac{G_F\mu_Z^2\sqrt{2}}{\hbar c}c_\mathrm{V}^ec_\mathrm{A}^e,
\end{align}
should be used for relevant interactions, where $G_F$ is the Fermi coupling constant, $\mu_Z=m_Zc/\hbar$, $m_Z=91.1880(20)$ GeV \cite{navas_review_2024}, $c^e_\mathrm{V}=-\frac{1}{2}(1-4\sin^2\theta_W)$, $c^e_\mathrm{A}=-\frac{1}{2}$, and $\sin^2\theta_W=0.23873(5)$ \cite{navas_review_2024}. In the contact limit approximation ($\mu\rightarrow\infty$), which is valid for $m_Z$, 
 \begin{align}
     \frac{e^{-\mu r_{jk}}}{r_{jk}} \rightarrow \frac{4\pi}{\mu^2}\delta^3(r_{jk}),
 \end{align}
 simplifying $V^{\mathrm{VA}}_Z$ to the more familiar form (e.g. Ref.~\cite{blundell_highaccuracy_1990, blundell_highaccuracy_1992})
 \begin{align}
     V^{\mathrm{VA}}_Z(\mu\rightarrow\infty) = \frac{G_F\sqrt{2}}{2}c_\mathrm{V}^ec_\mathrm{A}^e\sum_{j\neq k}\delta^3(r_{jk})(\gamma_j^5+\gamma_k^5). \label{eq:Vcontact}
 \end{align}
 
 Additionally, bounds on coupling constants $|g_\mathrm{A} g_\mathrm{V}|_X$ for a new, leptophilic vector boson $X$ can be extracted. These dark matter candidates have less stringent constraints than those from quark-interacting bosons \cite{buras_global_2021}, particilarly outside the low-mass region \cite{cong_spindependent_2025}, meriting their investigation. Computing $D^{\mathrm{VA}}_{wv}$ for different masses $m_X$ and comparing with existing deviations between theory and experiment,
 \begin{align}
     \Delta D^{\mathrm{V}\mathrm{A}}_{wv}\equiv \left|D_{wv}^{\mathrm{expt}}-(D^{\mathrm{V}\mathrm{A}}_{wv})^{eN}_Z-(D^{\mathrm{V}\mathrm{A}}_{wv})^{ee}_Z\right|+\delta D_{wv},
 \end{align} where $\delta D_{wv}$ contains all uncertainties added in quadrature, yields constraints on the $X$-mediated interaction's coupling constants:
\begin{align}
    |g^e_{\mathrm{V}}g^e_{\mathrm{A}}|_X\le \frac{\Delta D^{\mathrm{V}\mathrm{A}}_{wv}}{|D_{wv}^{\mathrm{V}\mathrm{A}}|^{ee}_X}.\label{eq:gg_VA}
\end{align}

\section{Calculations}\label{sec:calc}

We perform calculations using the program \textsc{ampsci} \cite{roberts_electricdipole_2023, caddell_accurate_2023} which begins by applying the DHF procedure to produce Slater-determinant wavefunctions of Dirac orbitals
\begin{align}
    \phi_{n\kappa m}(\v{r})&=\frac{1}{r}\begin{pmatrix}
        f_{n\kappa}(r)\Omega_{\kappa m}(\hat{r}) \\ ig_{n\kappa}(r)\Omega_{-\kappa,m}(\hat{r})
    \end{pmatrix},
\end{align}
where $n,\kappa,m$ are quantum numbers, $f,g$ are radial components and $\Omega_{\pm\kappa, m}$ are spherical spinors. In the derivations that follow it will also be convenient to consider the radial spinor $F_{n\kappa}$ of orbital $\phi_{n\kappa m}$,
\begin{align}
    F_{n\kappa}\equiv\begin{pmatrix}
        f_{n\kappa}(r)\\g_{n\kappa}(r)
    \end{pmatrix}.
\end{align}
Corrections accounting for higher-order electron correlations are included using the all-orders correlation potential method \cite{dzuba_screening_1988, dzuba_summation_1989, roberts_electricdipole_2023}, which by introduces an additional potential to the Hamiltonian to account for hole-particle interactions, screening of correlation effects by the core, and `self-energy chaining', to all-orders. Additionally, the divalent $^{199}$Hg system is calculated with the CI+MBPT \cite{dzuba_combination_1996}, which treats valence-valence interactions with CI and core-valence interactions with MBPT.

This yields atomic wavefunctions which can be used to calculate the effects of new interactions. To first order, these effects are determined by calculating matrix elements of their respective operators. In Sec.  \ref{sec:me} we provide details of this procedure for a general Yukawa potential [Eq. (\ref{eq:yuk})] in a DHF context, as well as limiting approximations. This is followed in Sec. \ref{sec:higher_order} by an approach for including many-body corrections to the new interaction with the TDHF method.

\subsection{Matrix elements}\label{sec:me}
Any effect induced by an $\alpha\beta$ interaction requires calculations of matrix elements $\matel{\Psi_i}{V_0^{\alpha\beta}}{\Psi_v}$, which, analogously to the DHF potential \cite{johnson_atomic_2007}, can be expressed for single-valence systems (where $\Psi_i$ and $\Psi_v$ differ by exactly one state) as a sum over two-body integrals of single-particle orbitals by following the Slater-Condon rules:
\begin{align}\label{eq:slater-condon}
    \matel{\Psi_i}{V_0^{\alpha\beta}}{\Psi_v} &= \sum_{a\neq v}^{N_c} (u^{\alpha\beta}_{iava}-u^{\alpha\beta}_{iaav}),
\end{align}
where the sum is over all $N_c$ core electrons $a=\{n_a,\kappa_a,m_a\}$ and $u_{abcd}$ are two-body integrals over electron orbitals,
\begin{align}
    u^{\alpha\beta}_{abcd}=\iint\d^3 \v{r}_1 \d^3 \v{r}_2\phi_a^\dag(\v{r}_1)\phi_b^\dag(\v{r}_2)u^{\alpha\beta}(r_{12})\phi_c(\v{r}_1)\phi_d(\v{r}_2),
\end{align}
where $u^{\alpha\beta}(r_{12})$ is the effective two-body electron operator for the interaction,
\begin{align}
    u^{\alpha\beta}(r_{12})=\frac{\hbar c}{4\pi}\frac{e^{-\mu r_{12}}}{r_{12}}\gamma^0_1\gamma^0_2\Gamma^{\alpha\beta}_{12}.
\end{align}
In the contact limit for the VA interaction, a Fierz identity relates the \textit{direct} and \textit{exchange} integrals: $u^{\mathrm{VA}}_{iava}=-u^{\mathrm{VA}}_{iaav}$. In this case we therefore only calculate the direct part where the spatial terms can be neglected \cite{blundell_highaccuracy_1990, blundell_highaccuracy_1992}.

The integrals ${u}_{iava}$ can be evaluated by expressing $V_0^{\alpha\beta}$ in terms of its action on a state:
\begin{align}\label{eq:V}
    V_0^{\alpha\beta}\phi_v&(\v{r}_1) = \frac{\hbar c}{4\pi}\sum_{a\neq v}^{N_c}\biggl(\int\d^3\v{r}_2 \phi_a^\dag(\v{r}_2)\frac{e^{-\mu r_{12}}}{r_{12}}\nonumber\\& \gamma^0_1\gamma^0_2\Gamma^{\alpha\beta}_{12}\bigl(\phi_v(\v{r}_1)\phi_a(\v{r}_2) - \phi_a(\v{r}_1)\phi_v({\v{r}_2})\bigr)\biggr).
\end{align}
By separating the angular and radial integrals, the potential can be expressed in terms of the radial spinor $F_v$ as $[V_0^{\alpha\beta}F_v(r)]_n$, which is itself a radial spinor with $\kappa=\kappa_n$ and is defined such that $\matel{\phi_n}{V_0^{\alpha\beta}}{\phi_v}=\int\d r F^\dag_n [V_0^{\alpha\beta} F_v]_n$. Integrating over angles and summing over magnetic quantum numbers by expanding $e^{-\mu r_{12}}/r_{12}$ over spherical harmonics \cite{winkelmann_solution_2021} yields
\begin{align}
    [V_0^{\alpha\beta}&F_v(r)]_n =\frac{\mu\hbar c\delta_{\kappa_n,\widetilde{\kappa}_v}}{4\pi}\Bigl(V_\mathrm{dir}^{\alpha\beta}(r)F_v(r) +[V_\mathrm{exch}^{\alpha\beta}F_v(r)]\Bigr),
\end{align}
where $\delta$ is a Kronecker delta, $\widetilde{\kappa}\equiv\mathcal{P}(\alpha\beta)\kappa$, and $\mathcal{P}(\alpha\beta)=\pm1$ is the parity of the interaction as given in Table~
\ref{tab:interactions}. 

The \textit{direct} contribution is
\begin{align}\label{eq:dir}
   V_\mathrm{dir}^{\alpha\beta}(r)F_v(r)=
       \sum_{n_a,\kappa_a}[j_a]B^{0,\beta\alpha}_{aa}(r)F_v(r)
\end{align}
for $\alpha\beta\neq\mathrm{PP,AA}$. In those cases there is no direct contribution due to the $\gamma^5$ terms, which, along with the mixing of $f$ and $g$ radial components in the exchange case, implies that these interactions are suppressed relative to their counterparts. Here, $j_a$ is the total angular momentum eigenvalue of $\phi_a$, $[x]=2x+1$, the sum is over all core orbitals $\{n_a,\kappa_a\}$ and the $B^{\lambda,\alpha\beta}_{ab}(r)$ terms are $2\times2$ matrices which encode the radial integrals and the Dirac matrices' effect on the radial spinors,
\begin{align}\label{eq:B}
    B^{\lambda,\alpha\beta}_{ab}(r)\equiv \int \mathrm{d}r' i_\lambda(\mu r_<)k_\lambda(\mu r_>)F^\dag_a(r') \mathcal{T}^{\alpha\beta}_{r,r'}F_b(r'),
\end{align}
where $i_\lambda$, $k_\lambda$ are modified spherical Bessel functions, $r_<=\min(r,r')$, $r_>=\max(r,r')$ and $\mathcal{T}^{\alpha\beta}_{r,r'}$ are combinations of generalised Pauli matrices which act on \textit{radial} spinors and are summarised in Table~\ref{tab:interactions}. To evaluate Eqs. (\ref{eq:dir}) and (\ref{eq:exch}) we observe that $\mathcal{T}^{\alpha\beta}_{jk}=\mathcal{T}^{\beta\alpha}_{kj}$.

The \textit{exchange} contribution is
\begin{align}\label{eq:exch}
    [V^{\alpha\beta}_\mathrm{exch}&F_v(r)]=-\sum_{n_a,\kappa_a}\frac{(-1)^{j_v-j_a}}{[j_v]}\sum_{\lambda=|j_v-j_a|}^{j_v+j_a}[\lambda]C^{\lambda, \alpha\beta}_{av}(\hat{r})\nonumber\\&\times\begin{cases}
        B^{\lambda,\alpha\beta}_{av}F_a(r),&\alpha=\beta,\\
        \left(B^{\lambda,\alpha\beta}_{av}(r)+B^{\lambda,\beta\alpha}_{av}(r)\right)F_a(r),&\alpha\neq\beta,
    \end{cases}
\end{align}
where
\begin{align}\label{eq:C}
    C&^{\lambda,\alpha\beta}_{av}\equiv\redmatel{\mathcal{P}(\alpha\beta)\kappa_v}{C^\lambda}{\mathcal{P}(\alpha)\kappa_a}\redmatel{\kappa_a}{C^\lambda}{\mathcal{P}(\beta)\kappa_v}
\end{align}
is a product of reduced matrix elements of the spherical tensor $C^\lambda_q(\hat{r})\equiv\sqrt{4\pi/[\lambda]}Y_{\lambda q}(\hat{r})$ with $Y_{\lambda q}$ the spherical harmonic, $\mathcal{P}(\mathrm{S})=\mathcal{P}(\mathrm{V})=+1$ and $\mathcal{P}(\mathrm{P})=\mathcal{P}(\mathrm{A})=-1$.

Recent SP calculations \cite{stadnik_improved_2018, maison_axionmediated_2021, maison_static_2022, prosnyak_updated_2023} have extracted bounds in the massless and contact limits,
\begin{align}
    i_\lambda(\mu r_<)k_\lambda(\mu r_>)=\begin{cases}
        r^\lambda_</(\mu[\lambda]r^{\lambda+1}_>),&\mu\rightarrow0,\\\delta(r-r')/(r^2\mu^3),&\mu\rightarrow\infty.
    \end{cases}
\end{align}
To validate our method and to determine for which masses the approximations can be safely applied, we also perform calculations using both limiting cases.

Evaluating effects arising from $\mathcal{P}$-odd interactions mixing  opposite-$\mathcal{P}$ states, i.e. Eq. (\ref{eq:wOv}), also requires calculating matrix elements of $T$. It can be convenient to express these in a similar manner to those of ${V}^{\alpha\beta}_0$, even if they are one-body operators. Here, we outline the process for the atomic electric dipole operator. For single-valence systems, its matrix elements are
\begin{align}
    \matel{\Psi_a}{\v{D}}{\Psi_b} = -|e|\matel{\phi_a}{\v{r}}{\phi_b}
\end{align}
as per the Slater-Condon rules. Using the spherical basis, where $r_q=rC^1_q$, separates the angular and radial integrals. Choosing the $z$-component, $q = 0$, and applying the Wigner-Eckart theorem, yields
\begin{align}
    \matel{\Psi_a}{\v{D}_z}{\Psi_b} = (-1)^{j_a-m_a}\begin{pmatrix}
        j_a & 1 & j_b \\ -m_a & 0 & m_b
    \end{pmatrix}\redmatel{a}{{d}}{b},
\end{align}
where the reduced matrix element is
\begin{align}
    \redmatel{a}{{d}}{b} &= \redmatel{\kappa_a}{C^1}{\kappa_b}\int_0^\infty \d r F_a^\dag r F_b.
\end{align}

\subsection{Many-body effects}\label{sec:higher_order}

Since DHF wavefunctions are self-consistently determined with all intra-atomic potentials, it seems natural to include $V_0^{\alpha\beta}$ in the Hamiltonian, as performed in Refs.~\cite{fleig_violating_2020, trantan_paritymixed_2022}, for example. However, for $\mathcal{P}$-odd potentials this requires breaking normal symmetry assumptions. Instead, many-body corrections due to the new potential can be found by self-consistently solving the Dirac equation with perturbed wavefunctions: the TDHF equations \cite{dirac_note_1930, dalgarno_timedependent_1966, johnson_relativistic_1980, dzuba_relativistic_1984a, dreuw_singlereference_2005}. Solving these yields first-order corrections to the DHF potential and the electron orbitals from the new interaction. In principle, higher-order effects could also be probed, however, as the new interactions are necessarily  smaller than Standard Model effects, higher order corrections from those will dominate.

In the presence of an external field $T^k_q$ (an irreducible tensor operator of rank $k$ and projection $q$), the single-particle orbitals $\phi$, DHF Hamiltonian $h$, and electron energies $\varepsilon$ are perturbed as
\begin{subequations}
    \begin{align}
        \phi&\rightarrow\phi+\delta\phi,\\
        h&\rightarrow h+\delta V_{\mathrm{HF}}+T^k_q,\\
        \varepsilon&\rightarrow\varepsilon+\delta\varepsilon.
    \end{align}
\end{subequations}
Substituting these into the Dirac equation and only keeping terms perturbed up to first-order relates the perturbations to the DHF orbitals,
\begin{align}\label{eq:TDHF_inv}
    (h-\varepsilon)\delta\phi = -(T^k_q+\delta V_\mathrm{HF}-\delta\varepsilon)\phi.
\end{align}
For frequency-independent fields (e.g.~$V_0^{\alpha\beta}$), solving Eq.~(\ref{eq:TDHF_inv}) self-consistently for $\delta\phi$ and $\delta V_{\mathrm{HF}}$ yields the first-order corrections due to $T^k_q$. The $\mathcal{P}$-odd case is further simplified because $\delta \varepsilon=0$.

If $T^k_q$ varies with frequency $\omega$, the oscillation can be separated by defining $T^k_q=t^k_q e^{-i\omega t}+(t^k_q)^\dag e^{i\omega t}$, where $t^k_q$ is an irreducible tensor operator of rank $k$ and projection $q$. The general perturbations become slightly more complex,
\begin{subequations}
\begin{align}
    \delta\phi&=Xe^{-i\omega t}+Ye^{i\omega t},\\
    \delta \varepsilon&\rightarrow\delta\varepsilon(e^{-i\omega t}+e^{i\omega t}),
\end{align}    
\end{subequations}
where $X$ and $Y$ are orbitals to be solved and, in the $\mathcal{P}$-odd case, $\delta\varepsilon$ remains zero. The TDHF equations then become
\begin{subequations}
\begin{align}
    (h_{\mathrm{HF}}-\varepsilon-\omega)X&=-(t^k_q+\delta V_{\mathrm{HF}}-\delta \varepsilon)\phi, \\
    (h_{\mathrm{HF}}-\varepsilon+\omega)Y&=-(t^k_q+\delta V_{\mathrm{HF}}^\dag-\delta \varepsilon)\phi.
\end{align}
\end{subequations}

\section{Results and Discussion}

We compute EDMs induced by $\mathcal{P}$-odd SP potentials for all mediator masses and $E$1 transitions due to the VA interaction in the contact limit. All results determine matrix elements $\matel{\widetilde{\Psi}_w}{\v{D}}{\widetilde{\Psi}_v}$ as per Eq. (\ref{eq:wOv}) and apply both TDHF and all-order correlation corrections, providing a robust set of bounds on new SP and VA interactions. These are then compared with experimental measurements to extract constraints on new physics.

\begin{figure}
    \centering
    \includegraphics[width=\linewidth]{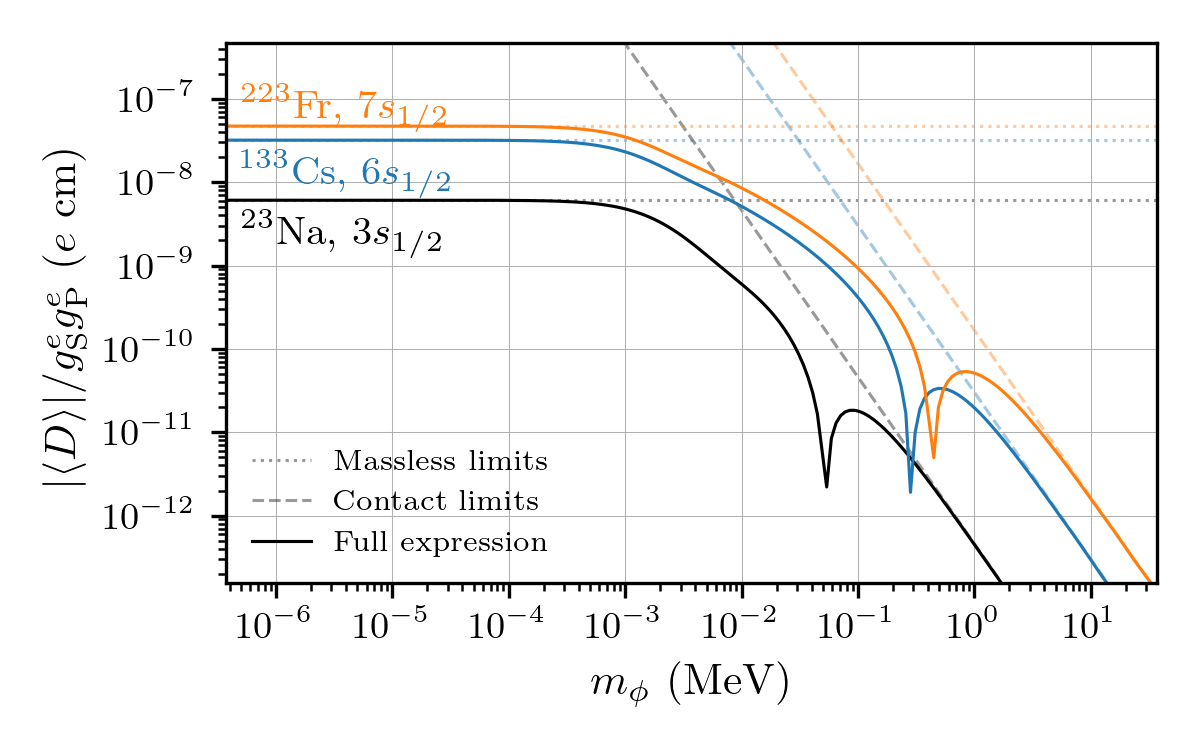}
    \caption{Atomic EDMs in the presence of an SP interaction ($|D^{\mathrm{SP}}_{v_0v_0}|$, ground state $v_0$) using the full Yukawa potential for all masses [Eqs.~(\ref{eq:yuk}), (\ref{eq:wOv}), (\ref{eq:slater-condon})--(\ref{eq:exch})] converge to the massless and contact limits (dotted and dashed lines). Troughs indicate sign changes in $D^{\rm SP}_{v_0,v_0}$ [see Fig.~\ref{fig:SP} (inset)] and produce regions of insensitivity in exclusion plots for the interaction.}
    \label{fig:limits}
\end{figure}

We begin by validating the intermediate mass approach with comparison to the limiting approximations. Figure~\ref{fig:limits} presents electric dipole matrix elements induced by SP interactions for a selection of single-valence systems, demonstrating convergence to the massless ($\mu\ll 1/a_0$ for Bohr radius $a_0$) and contact ($\mu\gg Z/a_0$) limits, where $D^{\alpha\beta}$ is $\mu$-independent in the former case and $\propto1/\mu^2$ in the latter. A sign change, also seen in Refs.~\cite{stadnik_improved_2018, maison_axionmediated_2021, maison_static_2022, prosnyak_updated_2023}, occurs in the intermediate range and manifests in regions of insensitivity to experimental constraints.

\subsection{Scalar-pseudoscalar}
\begin{figure}
    \centering
    \includegraphics[width=\linewidth]{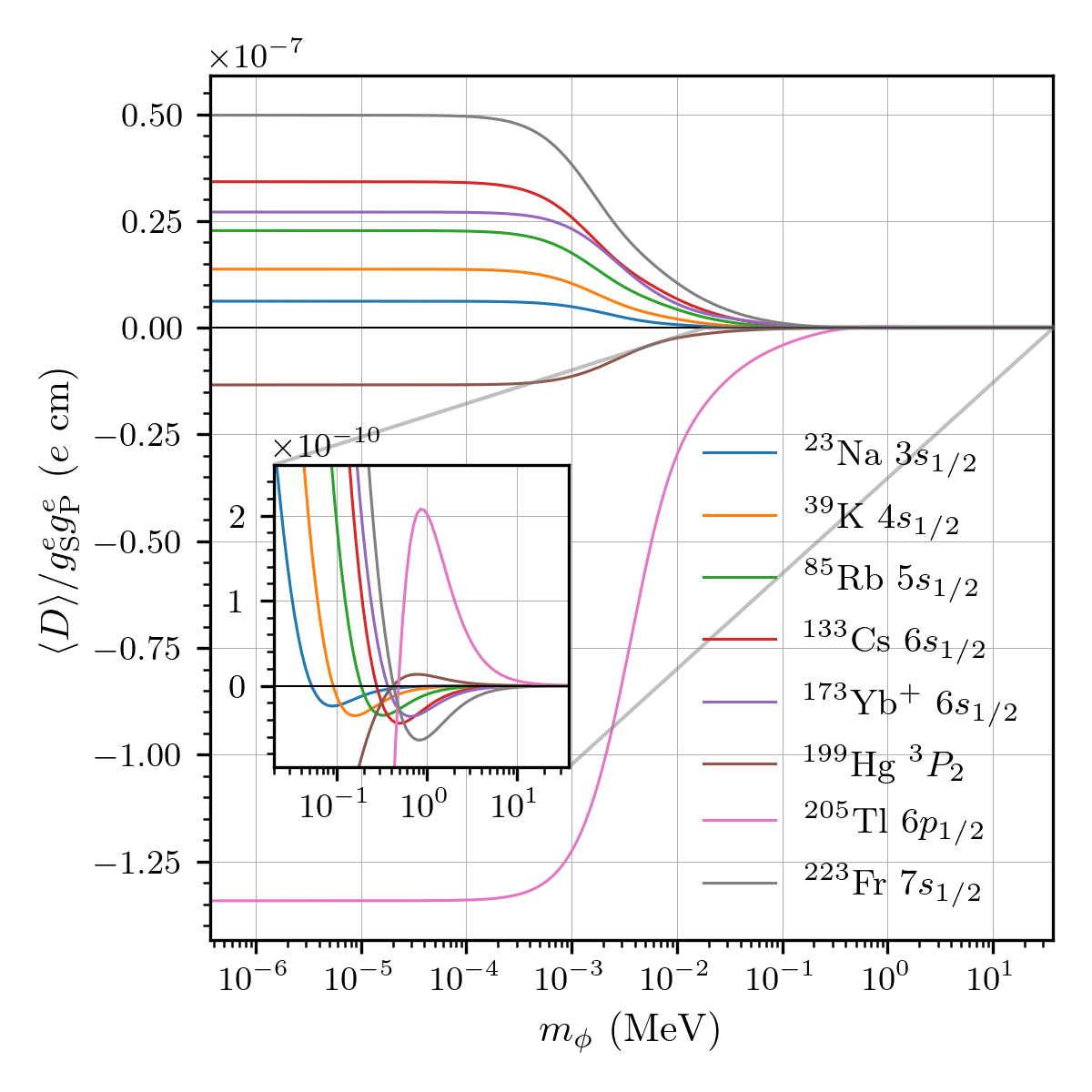}
    \caption{Calculations of ground state atomic EDMs induced by an SP interaction, $D^{\mathrm{SP}}_{v_0v_0}$, for a selection of atoms. EDMs are stronger for larger systems and change sign when the mediator's mass $m_\phi$ is near the electron mass $m_\phi\approx m_e$.}
    \label{fig:SP}
\end{figure}

Atomic EDMs for a range of atoms are presented in Fig.~\ref{fig:SP}, each treated as a single-valence system (except divalent $^{199}$Hg). Their $Z$-dependence \cite{stadnik_improved_2018} is clearest for group I elements where configuration-specific contributions are similar and indicates that heavier systems are generally most affected by new electron-electron interactions. Nevertheless, the spread of EDM sensitivity across systems (approximately a single order of magnitude) is small compared to variation in existing experimental observations \cite{roberts_parity_2015}, so future measurements should preference systems with lower experimental uncertainty over those with larger EDM sensitivity.

The $^{205}$Tl calculations are an order of magnitude larger than the rest due to a larger TDHF contribution. This system differs from the others by its trivalent configuration and ground state angular momentum ($p_{1/2}$). Interactions between the tight valence electrons are the likely source of many-body enhancements arising at the TDHF level and should be investigated in future work by comparison with, for example, the CI+MBPT method (as was used in Ref.~\cite{stadnik_improved_2018}).

\begin{figure}
    \centering\includegraphics[width=\linewidth]{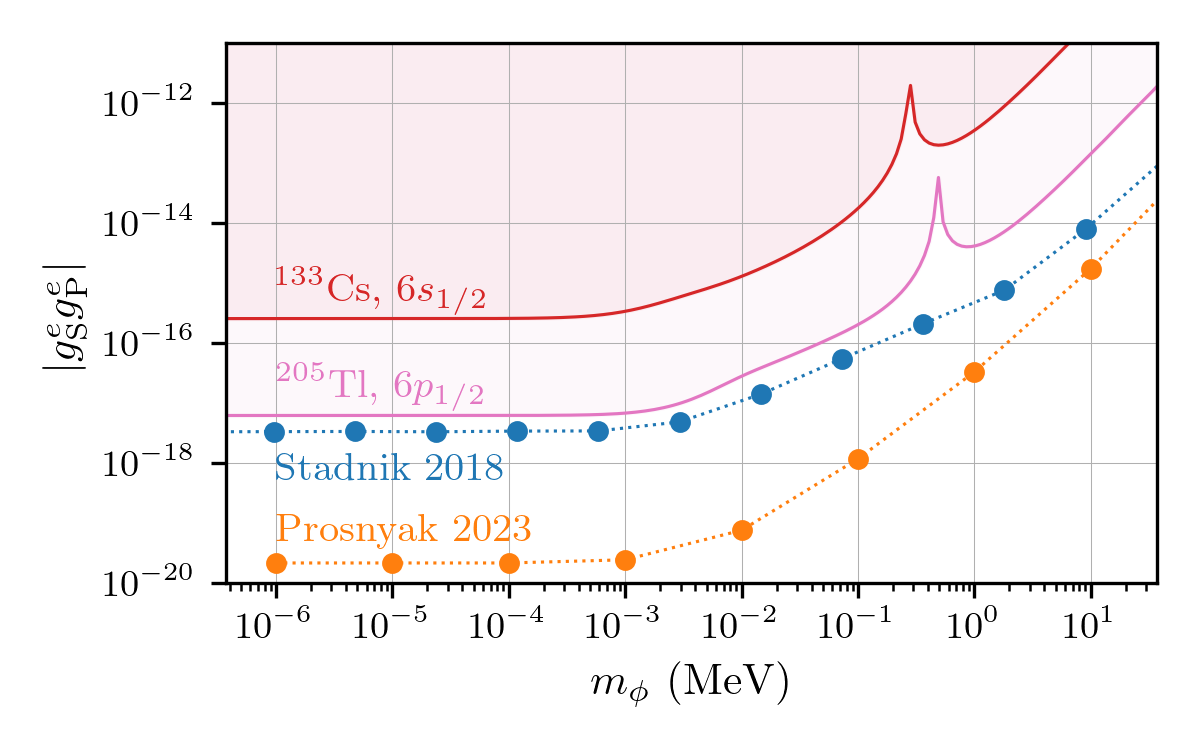}
    \caption{Upper bounds on the coupling constants of a new SP electron-electron interaction determined by comparing calculations (Fig.~\ref{fig:SP}) with experimental atomic EDM limits. Existing bounds compiled from multiple systems \cite{stadnik_improved_2018} and from HfF$^+$ \cite{prosnyak_updated_2023} are also shown.}
    \label{fig:SP_bounds}
\end{figure}

The EDM calculation for $^{205}$Tl, when combined with its most precise measurement  ($-0.40(43)\times10^{-24}$ $e$ cm \cite{regan_new_2002}), yields the strictest upper bound on the SP coupling strength from paramagnetic atoms, as shown in Fig.~\ref{fig:SP_bounds}. The bounds are similar to those obtained in Ref.~\cite{stadnik_improved_2018} from molecular and diamagnetic systems, which were then surpassed \cite{prosnyak_updated_2023} due to updated experimental bounds \cite{roussy_improved_2023}. Nevertheless, the molecular measurements are considerably more recent than the atomic: the $^{205}$Tl observation was made in 2002, while the molecular in 2023. We hope that our method and extracted bounds renew interest in \textit{atomic} electric dipole moments for paramagnetic systems as they too are viable probes of new physics.

We also present extracted bounds for $^{133}$Cs which has the next most stringent paramagnetic bound ($-0.18(69)\times10^{-23}$ $e$ cm \cite{murthy_new_1989}). The difference from $^{205}$Tl (about 100 times less stringent) is in similar proportions due to the experiments' uncertainties and the many-body enhancement in thallium. Only by including many-body external field corrections to thallium does the enhancement become evident; future calculations should include such effects for robust results. 

The strictest bound on atomic EDMs is from $^{199}$Hg: $-2.20(2.75)_{\rm stat}(1.48)_{\rm syst}\times10^{-30}$ $e$ cm \cite{graner_reduced_2016a}. Selection rules imply that neutral mercury admits no ground state $ee$-induced EDM by mixing $V_0^{\rm SP}$ with $\v{D}$, so Eq.~(\ref{eq:wOv}) cannot be directly used to extract a new physics constraint. However, an EDM may be obtained by mixing with third operator, such as the hyperfine interaction, at second order in perturbation theory. This was investigated in Ref.~\cite{stadnik_improved_2018} using calculations for the Standard Model weak interaction \cite{flambaum_new_1985, martensson-pendrill_calculations_1987, dzuba_relations_2011} to obtain contact limit constraints, but can in principle be extended via Eqs.~(\ref{eq:slater-condon})--(\ref{eq:exch}) to an arbitrary scalar boson. Similar investigations have also been performed and discussed in Refs.~\cite{fleig_modelindependent_2018, hubert_electric_2022} for electron-nuclear interactions and electron EDMs.

Finally, in light of the precision in the experimental ground state EDM in mercury, we also consider its metastable~$^3P_2$ state, which we calculate to have a lifetime of $0.68$s. Assuming that such a lifetime could give rise to an EDM experiment with similar precision to Ref.~\cite{griffith_improved_2009}, Fig.~\ref{fig:SP} demonstrates that accurate calculations with many-body external field corrections for divalent systems can also be performed to extract new bounds.

\subsection{Vector-axial vector}

\begin{table}
    \centering
    \caption{Electron-electron neutral weak contribution to the $6s\rightarrow7s$ $E_{\rm PNC}$ transition amplitude in $^{133}$Cs. Results provided at the Dirac-Hartree-Fock (DHF) level and with all-order correlation potential corrections ($\Sigma^\infty$) \cite{dzuba_relativistic_1984, dzuba_relativistic_1985, dzuba_screening_1988, dzuba_summation_1989} and first-order corrections due to the $Z^0$-induced interaction (TDHF), see Eq.~(\ref{eq:TDHF_inv}). The previous calculation \cite{blundell_highaccuracy_1990, blundell_highaccuracy_1992} is updated with $\sin^2\theta_W=0.23873(5)$.\cite{navas_review_2024} for comparison.}
    \begin{ruledtabular}
    \begin{tabular}{lll}
    Source & Contribution & $E_{\rm PNC}^{ee}$ ($i|e|a_0$) $\times10^{-11}$ \\
    \hline 
    Refs. \cite{blundell_highaccuracy_1990, blundell_highaccuracy_1992} & DHF & $-0.00018$\\ 
         This work & DHF & $-0.00019$ \\
         This work & +$\Sigma^\infty$+TDHF & $-0.00025$\\
    \end{tabular}
    \end{ruledtabular}
    \label{tab:Epnc_ee}
\end{table}

For the VA case, we first report calculations of the leptonic contribution to the $6s\rightarrow7s$ $^{133}$Cs transition amplitude $E^{ee}_{\mathrm{PNC}}$  in Table~\ref{tab:Epnc_ee}. Using the contact limit potential with neutral weak couplings [Eq. (\ref{eq:Vcontact})], the results agree with previous calculations at the DHF level. The inclusion of many-body corrections amount to 24\% of the final value, $E^{ee}_{\rm PNC}=-0.00025\times10^{-11}\,i|e|a_0$.

\begin{figure}
    \centering
    \includegraphics[width=\linewidth]{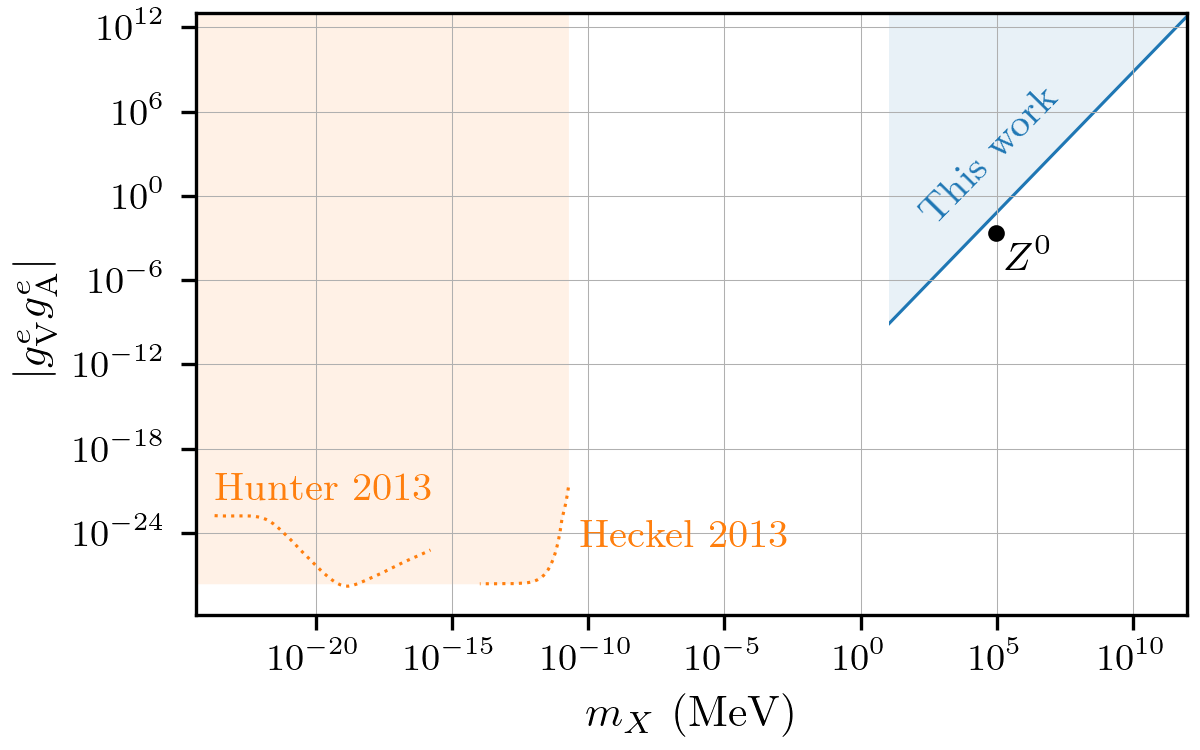}
    \caption{Upper bounds on the interaction strength of a new leptophilic vector boson in the contact limit, found by comparing the $^{133}$Cs $E_{\mathrm{PNC}}$ measurement \cite{wood_measurement_1997} and calculations \cite{dzuba_highprecision_2002, porsev_precision_2009, dzuba_revisiting_2012}  with a new boson's electron-electron contribution. The line thickness covers the spread arising from disagreeing vector transition polarizabilities $\beta$ \cite{dzuba_offdiagonal_2000, toh_determination_2019, trantan_reevaluation_2023}. Torsion pendulum experiments  \cite{heckel_limits_2013,hunter_using_2013} provide stringent low mass bounds. The $Z^0$ boson's mass and coupling are indicated for reference.}
    \label{fig:VA_bounds}
\end{figure}

In Fig.~\ref{fig:VA_bounds} we extract new bounds on an arbitrary vector boson in the contact limit by comparing existing $E^{eN}_{\mathrm{PNC}}$ calculations with experiment. As the experimental result depends on the vector transition polarisability $\beta$, this deviation was determined by
\begin{align}
    \Delta D^{\mathrm{VA}}_{6s,7s} = \left|\beta\left(\frac{-\mathrm{Im}(E_{\mathrm{PNC}})}{\beta}\right)_\mathrm{expt}-E_\mathrm{PNC}^\mathrm{th}\right| + \delta,
\end{align}
where $\delta$ contains propagated uncertainties. For the theory result, we use a nuclear weak charge of $Q_W=-73.26(1)$ \cite{navas_review_2024} to find $E_{\mathrm{PNC}}^{\mathrm{th}}=-0.8364(38)\times 10^{-11} i|e|a_0$  from previous high accuracy atomic calculations \cite{dzuba_highprecision_2002, porsev_precision_2009, dzuba_revisiting_2012}, and note that our calculated $ee$ contribution is smaller than this uncertainty. While $-\mathrm{Im}(E_{\mathrm{PNC}})/{\beta}=-1.5935(56)$ mV/cm \cite{wood_measurement_1997} is the primary experimental result, various determinations of $\beta$ exist which disagree. Considering $\beta=26.957(51)a_0^3$ \cite{dzuba_offdiagonal_2000}, $\beta=27.139(42)a_0^3$ \cite{toh_determination_2019} and $\beta=26.887(38)a_0^3$ \cite{trantan_reevaluation_2023} yields final deviations $\Delta D^{\mathrm{VA}}_{6s,7s}=0.0128,0.0071,$ and $0.0148$ respectively. This spread is covered by the thickness of the line in Fig.~\ref{fig:VA_bounds}. We note that a recent fermion-loop correction to $Q_W$ \cite{flambaum_effects_2026a}, extending those developed in Refs.~\cite{ghosh_neutrino_2025,ghosh_neutrino_2025a}, reduces the first and last discrepancies but increases the second, thereby leaving the spread unaffected.

These are the first atomic bounds for a leptonic VA interaction and the first direct probe of the high-mass sector (see Ref.~\cite{cong_spindependent_2025} for a discussion of combined astrophysical-laboratory limits). The contact limit exhibits explicit mass dependence, $|g^e_\mathrm{V}g^e_\mathrm{A}|\propto m_X^2$. This yields an extracted bound, for $m_X\gtrsim 10\,{\rm MeV}$,  of
\begin{align}
    |g^e_{\mathrm{V}}g^e_{\mathrm{A}}|/m_X^2 \le 10^{-11} \,\mathrm{MeV^{-2}}.
\end{align}

\vspace{-1em}
\subsection{Outlook}
Our next priority is to provide a treatment of the spatial contribution to the VA interaction to extract bounds in the currently unconstrained intermediate mass region. Noting that they must converge to a mass-independent expression in the low mass limit, we expect the intermediate and massless bounds to be at most few orders of magnitude more stringent than the strictest given by the contact limit. While unlikely to improve on low mass spin-torsion constraints, they will produce new intermediate-mass bounds. Alternative experiments (e.g.,\ Josephson junction interferometry \cite{croon_searching_2026} or magnetic force sensing \cite{amaral_levitatedmagnet_2026}) may complement these in the future.

$\mathcal{P}$-even interactions, also calculable with our method, should also briefly be mentioned. Light atoms may be most sensitive to detecting energy shifts induced by such an interaction because many-body effects are minimised. Since the most recent review \cite{cong_spindependent_2025}, these have been investigated in helium \cite{cong_testing_2026} with a large discrepancy in the $1s2s\,^3S_1$ ionisation energies of $^{3,4}$He \cite{patkos_complete_2021,clausen_ionization_2025, clausen_ionization_2025}, and in Li-like systems \cite{abdullin_axionexchange_2026}, which admit accurate QED treatments unlike the heavy systems considered here. Furthermore, heavier systems may be probed for $\mathcal{P}$-even interactions by considering, e.g., isotope shifts. These isotopic differences in transition frequencies have been a subject of renewed interest due their sensitivity to new physics \cite{berengut_precision_2025}.

\section{Conclusion}

We have presented a comprehensive method for calculating the atomic effects of any scalar or vector electron-electron interaction and extracted new physics bounds. Our approach enables investigation of the intermediate mass range where neither the massless nor contact limits apply and includes many-body corrections induced by the interaction for both $\mathcal{P}$-even and $\mathcal{P}$-odd potentials. It can be used to study physics beyond the Standard Model as well as the electron-electron weak interaction.

Calculations using this method were performed for $\mathcal{P}$-odd interactions of atomic EDMs (for the SP case) and transitions (for the VA case) to extract constraints on the interactions' coupling constants by comparison with experiment. While the SP results are in line with previous constraints, the VA limits provide new bounds for previously unconstrained heavy mediator masses. Our calculations also update the neutral weak electron-electron contribution to the $6s\rightarrow7s$ transition in Cs. Constraints on new physics are highly dependent on experimental uncertainties and we hope that these results spur renewed interest in atomic electric dipole moments and transitions as probes of physics beyond the standard model.

\acknowledgments
\noindent
This work was supported by the UQ Fellowship of the Big Questions Institute, by Australian Research Council through DECRA DE210101026 and Discovery Project DP230101685, by the Commonwealth through an Australian Government Research Training Program Scholarship [DOI: https://doi.org/10.82133/C42F-K220] and by computational resources provided by The University of Queensland via the Friday supercomputer.

%

\end{document}